\documentclass[12pt, preprint]{aastex}

\newcommand{\kmsMpc}{\ensuremath{\mathrm{km\ s}^{-1}\mathrm{\ Mpc}^{-1}}}
\newcommand{\pl} {\textsc{PixeLens}} 
\newcommand{\HC} {\ensuremath{H_0}}
\newcommand{\HCi}{\ensuremath{H_0^{-1}}}
\newcommand{\LCDMHC}{\ensuremath{\HCi=14 \mathrm{\ Gyr}}}
\newcommand{\OldResult} {\ensuremath{\HCi=13.5^{+2.5}_{-1.3} \mathrm{\ Gyr}}}
\newcommand{\NewResult} {\ensuremath{\HCi=13.7^{+1.8}_{-1.0} \mathrm{\ Gyr}}}
\newcommand{\NewResultLocal} {\ensuremath{\HC=71^{+6}_{-8}\ \kmsMpc }}
\newcommand{\SimulationResult} {\ensuremath{\HCi=13.3^{+1.4}_{-0.6} \mathrm{\ Gyr}}}
\newcommand{\FakeResult} {\ensuremath{\HCi=13.7^{+1.5}_{-1.4} \mathrm{\ Gyr}}}
\newcommand{\appref}[1] {Appendix~\ref{#1}}

\newcommand{\secref}[1] {Section~\ref{#1}}
\newcommand{\figref}[1] {Figure~\ref{#1}}
\newcommand{\eqnref}[1] {Equation~\ref{#1}}

\newcommand{\FortyFiveDeg}{\ensuremath{45^\circ}}
\newcommand{\AlgorithmS}{Algorithm~S}
\newtheorem{constraint}{Data Constraint}
\newtheorem{prior}{Prior}
\newcommand{\figplace} {}

\title{A New Estimate of the Hubble Time with Improved Modeling of Gravitational Lenses}
\author{Jonathan Coles}

\begin{document}
\bibliographystyle{apj}

%%%%%%%%%%%%%%%%%%%%%%%%%%%%%%%%%%%%%%%%%%%%%%%%%%%%%%%%%%%%%%%%%%%%%%%%%%%%%%
\begin{abstract} 
%%%%%%%%%%%%%%%%%%%%%%%%%%%%%%%%%%%%%%%%%%%%%%%%%%%%%%%%%%%%%%%%%%%%%%%%%%%%%%

This paper examines free-form modeling of gravitational lenses using Bayesian
ensembles of pixelated mass maps. The priors and algorithms from previous work
are clarified and significant technical improvements are made.  Lens
reconstruction and Hubble Time recovery are tested using mock data from simple
analytic models and recent galaxy-formation simulations.  Finally, using
published data, the Hubble Time is inferred through the simultaneous
reconstruction of eleven time-delay lenses.  The result is \NewResult\
(\NewResultLocal).

\end{abstract}

\keywords{gravitational lensing; cosmological parameters}

%%%%%%%%%%%%%%%%%%%%%%%%%%%%%%%%%%%%%%%%%%%%%%%%%%%%%%%%%%%%%%%%%%%%%%%%%%%%%%
\section{Introduction} \label{sec:Introduction}
%%%%%%%%%%%%%%%%%%%%%%%%%%%%%%%%%%%%%%%%%%%%%%%%%%%%%%%%%%%%%%%%%%%%%%%%%%%%%

Gravitational lenses provide a fantastic natural tool for probing many of the
large scale properties of the cosmos. Recent applications range from
estimating the age of the Universe \citep{2006ApJ...650L..17S} to studying the
dark matter profiles of galaxies \citep{2007arXiv0704.3267R} to testing
alternative theories of gravity \citep{2007arXiv0709.3189F}.

Despite their potential, gravitational lenses (GLs) are difficult to study
because of several degeneracies such as the position of the source and the mass
distribution of the lensing object. This paper focuses on strong lensing of
quasars by galaxies, but the techniques developed can equally be applied to
clusters.  Many have tried to fit models to GLs by assuming different galaxy
structures.  \cite{1981ApJ...244..736Y} were the first to do so with King
models and many others have followed using a variety of single isothermal
ellipses (SIEs), S\`ersic models, or de Vaucouleurs profiles \citep[for a
review see][]{2004astro.ph..7232K}.  But different models can easily give
different results \citep{2007A&A...464..845V}.

This kind of modeling is generally called parametric modeling. Each model has a
nominal amount of parameters that can be adjusted. But while one model may fit
the data, the degeneracies make it difficult to determine how well these models
really represent the lens; and as pointed out by \cite{1999AJ....118...14B} and
more recently by \cite{2007arXiv0704.3267R}, in connection with time delays,
without extreme care these models can be very sensitive to the assumptions. 

In contrast, free-form or non-parametric models reconstruct the lens on a grid
or a set of basis functions. No particular form is assumed and the results
allow a wider range of solutions than parametric models might. Such modeling is
not unique to lensing, though.

\cite{1979ApJ...232..236S} used non-parametric modeling to show for the first
time that it is possible to construct a triaxial stellar system in equilibrium.
He showed that there existed a distribution of stars on orbits that fit a given
density function $D$. The three-dimensional space of a galaxy was divided into
$M$ cells and $D$ was expressed by $D(J) = \sum_{I=1}^M C(I) \cdot B(I,J)$,
where $B(I,J)$ is the orbit density for an orbit $I$ in cell $J$, calculated
using test particles in a fixed potential. $C(I)$, the number of stars on orbit
$I$, was determined numerically by solving a linear program.

%To begin answering this the
%three-dimensional space of a galaxy was divided into $M$ cells and the orbits
%of test particles were calculated assuming a fixed potential. This resulted in
%an orbit density $B(I,J)$ for an orbit $I$ in cell $J$. With the orbit density
%the only thing left to find is the number of stars on each orbit such that
%$D(J) = \sum_{I=1}^M C(I) \cdot B(I,J)$. Since this equation is linear and
%obviously $C(I) \ge 0$ Schwarzschild was able to use standard linear
%programming techniques to find a solution numerically.

In a very similar manner, Schwarzschild's technique can be applied to lenses.
Modeling the lenses on a grid was first introduced by
\cite{1997MNRAS.292..148S} and then later extended to include both weak and
strong lensing by \cite{1998AJ....116.1541A}. Similar methods have also been
used by \cite{2005MNRAS.360..477D} and \cite{2005A&A...437...39B}.  But in
contrast to Schwarzschild, it is desirable to show the \emph{variety} of
solutions rather than just existence.  This important feature is incorporated
into the work of \cite{2000AJ....119..439W} and the software \pl\
\citep{2004AJ....127.2604S} (see \appref{AppendixA}). Related approaches are developed in
\cite{2000ApJ...535..671T} and \cite{2003ApJ...590...39K}.  Given a large
ensemble of models, one or several variables are examined while averaging out
(marginalizing) the others. The same principle is used in statistical
mechanics.  However, the use of marginalization is sometimes overlooked,
leading to a misunderstanding that pixelated models are ``grossly
underconstrained'' because the number of variables exceeds the number of data
points.

Pixelated modeling has the advantage of allowing the form of the lens to vary.
It does not presuppose important parameters and can produce models that would
otherwise not be possible with parametric modeling.  For instance, while
parametric models already showed that steepness is an important parameter
\citep{1994AJ....108.1156W}, pixelated models showed that shape degeneracies,
which are often difficult to capture with parametric models, cannot be ignored
\citep{2006ApJ...653..936S}; twists and nonuniform stretching are also easily
found.

In this paper, pixelated lens modeling and the constraints imposed on the
models are explicitly defined. The algorithms are improved with several
optimization techniques and the enhanced method is tested against lenses from
an $N$-body simulation and another fictitious data set. Finally, a system of
eleven lenses is used in the same way as \cite{2006ApJ...650L..17S} to further
constrain the Hubble Time.

%%%%%%%%%%%%%%%%%%%%%%%%%%%%%%%%%%%%%%%%%%%%%%%%%%%%%%%%%%%%%%%%%%%%%%%%%%%%%%
\section{Creating Models} \label{sec:Creating Models}
%%%%%%%%%%%%%%%%%%%%%%%%%%%%%%%%%%%%%%%%%%%%%%%%%%%%%%%%%%%%%%%%%%%%%%%%%%%%%

\pl\ generates an ensemble of lens models that fit the input data. In the
Bayesian way, the ensemble itself provides estimates and uncertainties.  Each
model consists of a set of $n$ discrete mass squares with density $\kappa_n$, a
source position $\vec\beta$, and optionally, a variable $h$ which is
proportional to \HC. If the time delays are unknown, the value of $h$ is fixed.
In this paper, where the time delays are known, $h$ varies across the ensemble.
The positions of observed images and the redshifts of the source and lens are
taken to be given with errors small enough to be ignored.  Time delays between
images, when available, are similarly assumed to be accurate.  Tests from
\cite{2006ApJ...650L..17S} show that adjusting these numbers slightly to
simulate errors has much less effect than the model uncertainties.  

The mass density in each square, or \emph{pixel}, is the projected mass density
on the plane of the sky in units of the critical density.%
\footnote{Many have suggested that it would be better to discretize the
potential, but the potential is not naturally discrete and doing so would
require recovering the mass from Poisson's equation; guaranteeing that the mass
remains positive is difficult and involves a double derivative which
produces noisy results.} The pixelated surface is a disc of radius {\tt pixrad}
pixels.  The total number of pixels is then about $\pi\cdot\mathrm{\tt
pixrad}^2$.  The extent of the modeled mass, {\tt maprad}, defaults to
\mbox{$\min\{{r_\mathrm{max} + r_\mathrm{min}}, 2r_\mathrm{max} -
r_\mathrm{min}\}$}, where $r_\mathrm{min}$ and $r_\mathrm{max}$ are the
distances of the innermost and outermost images, respectively. This allows for
a buffer zone outside the outermost image when required.

Following \cite{1986ApJ...310..568B}, the arrival time is the light travel time scaled by 
\begin{equation}
h^{-1}T(z_L, z_S) = (1 + z_L)\frac{D_L D_S}{cD_{LS}}
\end{equation}
where $z_L$ is the redshift of the lens, and $D_L, D_S$, and $D_{LS}$ are the
distances from observer to lens, observer to source, and lens to source,
respectively. This removes the dependence on a particular cosmology. The
$h^{-1}$ factor comes through the distance factors.

The arrival time at position $\vec\theta$ is given by
\begin{equation}
\tau(\vec\theta) = \frac{1}{2}|\vec\theta|^2%
                 - \vec\theta \cdot \vec\beta%
                 - \int \ln|\vec\theta - \vec\theta'|\kappa(\vec\theta')d^2\vec\theta'.
\label{eqn:lens}
\end{equation}
This can be interpreted as a surface, which is modeled with a summation over the pixel densities,
\begin{equation}
\begin{array}{ccl}
\tau(\vec\theta) &=& \frac{1}{2}|\vec\theta|^2%
                              - \vec\theta \cdot \vec\beta%
                              - \sum_n\kappa_n Q_n(\vec\theta) \\%
                             &+& \gamma_1(\theta_x^2 - \theta_y^2) + 2\gamma_2\theta_x\theta_y.
\end{array}
\label{eqn:lens-discrete}
\end{equation}
Two additional terms involving $\gamma_1$ and $\gamma_2$ are added to account
for external shear from neighboring galaxies.

The function $Q$ is the integral from (\ref{eqn:lens}) evaluated over a square
pixel with side length $a$.  $Q$ is defined using the same notation as in
\cite{1997MNRAS.292..148S}: Let $x,y$ be the Cartesian components of
$\vec\theta$, $r^2 = x^2 + y^2$ and 
\begin{equation}
\begin{array}{lcl}
\tilde{Q}_n(x,y) =  (2\pi)^{-1}[ & & x^2\arctan(y/x) \\
                                 &+& y^2\arctan(x/y) \\
                                 &+& xy(\ln r^2 - 3)]
\end{array}
\end{equation}
Then
\begin{equation}
\begin{array}{lcl}
Q_n(x,y)        &=& \tilde{Q}_n(x_+,y_+) 
                + \tilde{Q}_n(x_-,y_-) \\
                &-& \tilde{Q}_n(x_-,y_+) 
                - \tilde{Q}_n(x_+,y_-),
\end{array}
\end{equation}
where $x_{\pm} = x - x_n \pm a/2$ and $y_{\pm} = y - y_n \pm a/2$.

The function $\tau$ is linear in all the unknowns $\vec\beta, \kappa_n,
\gamma_1, \gamma_2$.  Constraints are placed on $\tau$ and the unknowns so that
the results are physical.  The data constraints come directly from lensing
theory.  The priors, or assumptions, are additional constraints that are
physically motivated.

%%%%%%%%%%%%%%%%%%%%%%%%%%%%%%%%%%%%%%%%%%%%%%%%%%%%%%%%%%%%%%%%%%%%%%%%%%%%%

As a side note, the source position can be negative because it is relative
to the center, but it must be positive in order to encode it as part of the linear program.
This is resolved by adding a constant internally.

%%%%%%%%%%%%%%%%%%%%%%%%%%%%%%%%%%%%%%%%%%%%%%%%%%%%%%%%%%%%%%%%%%%%%%%%%%%%%
\begin{constraint} Images are observed where the arrival time surface
is stationary, \mbox{$\vec\nabla\tau(\vec{\theta_i}) = 0$} (Fermat's Principle). 

\begin{equation}
\begin{array}{l}
\vec\theta_{i,x} - \vec\beta_x - \sum dQ/d\vec\theta_{i,x} = 0, \\
\vec\theta_{i,y} - \vec\beta_y - \sum dQ/d\vec\theta_{i,y} = 0,
\end{array}
\end{equation}
\label{con:grad}
\end{constraint}

%%%%%%%%%%%%%%%%%%%%%%%%%%%%%%%%%%%%%%%%%%%%%%%%%%%%%%%%%%%%%%%%%%%%%%%%%%%%%
\begin{constraint} 
The time delay between two images $\vec\theta_i$ and $\vec\theta_j$ must be
consistent with observations,
\begin{equation}
\tau(\vec{\theta_i}) - \tau(\vec{\theta_j}) = h\frac{[\mathrm{obs\ delay}]}{T(z_L, z_S)}.
\end{equation}
If the time delays are unknown the time ordering can be inferred from the
morphology and imposed by 
\begin{equation}
\tau(\vec{\theta_i}) - \tau(\vec{\theta_j}) \ge 0.
\end{equation}
\label{con:time 1}

%The dimensionless number $g$ is defined such that \[\HCi = g\mathrm{\ Gyr},\] where $g=9.78/h$.

\end{constraint}

%%%%%%%%%%%%%%%%%%%%%%%%%%%%%%%%%%%%%%%%%%%%%%%%%%%%%%%%%%%%%%%%%%%%%%%%%%%%%
\begin{constraint} \mbox{}
At each $\theta_i$ there are two constraints of the form
\begin{equation}
\epsilon\biggl|\frac{\partial^2}{\partial\theta_{x'}^2}\tau(\vec{\theta_i})\biggr|
\le
\biggl|\frac{\partial^2}{\partial\theta_{y'}^2}\tau(\vec{\theta_i})\biggr|
\end{equation}
where $\theta_{x'}$ and $\theta_{y'}$ are the local radial and tangential directions and
$\epsilon = 1/10$ by default.
\end{constraint}
This ensures that the image elongation is between $\epsilon$ and $1/\epsilon$ when
projected along the radial direction. In practice, the default does not place any
constraints on the image. If an image is known to be elongated then $\epsilon$
can be changed. In particular, this was used in \cite{1998AJ....116.1541A}.

%%%%%%%%%%%%%%%%%%%%%%%%%%%%%%%%%%%%%%%%%%%%%%%%%%%%%%%%%%%%%%%%%%%%%%%%%%%%%
\begin{constraint} If a model contains $N$ lenses, they must share the same Hubble Constant.
\begin{equation}
h_{\mathrm{lens}_1} = h_{\mathrm{lens}_2} = \dots = h_{\mathrm{lens}_N}
\end{equation}
\end{constraint}
When \HC\ is unspecified then \HC\ is allowed to vary from model to model
but not from lens to lens within a single model.

%%%%%%%%%%%%%%%%%%%%%%%%%%%%%%%%%%%%%%%%%%%%%%%%%%%%%%%%%%%%%%%%%%%%%%%%%%%%%
The following priors are the assumptions made about the lensing systems. All
are well-defined and astro-physically justified, as explained below. 

%%%%%%%%%%%%%%%%%%%%%%%%%%%%%%%%%%%%%%%%%%%%%%%%%%%%%%%%%%%%%%%%%%%%%%%%%%%%%
\begin{prior} The density cannot be negative.
\label{con:kappa positive}
\begin{equation}
\kappa_n \ge 0
\end{equation} 
\end{prior}
This is a quite trivial requirement, but one that can often be difficult to
ensure with other techniques. The linear programming algorithm employed here
guarantees this prior by design.

Notice the similarity between Schwarzschild's equation from the introduction on
the one hand and \eqnref{eqn:lens-discrete} and Prior~\ref{con:kappa positive}.
There is a linear function ($D$ or $\tau$) whose value is known, and a
summation over a product where one of the product terms is calculated
beforehand ($B$ or $Q$). Schwarzschild was limited at the time to what he could
say about the unknowns, but negative values were not allowed. The goal was only
to show the existence of one solution because no one knew at the time whether a
triaxial solution was possible. With lenses much more can be said about the
unknowns and lensing is known to occur.

%%%%%%%%%%%%%%%%%%%%%%%%%%%%%%%%%%%%%%%%%%%%%%%%%%%%%%%%%%%%%%%%%%%%%%%%%%%%%
\begin{prior} 
Most lens are assumed to have inversion symmetry, unless the lenses are observed
to be interacting or otherwise strongly asymmetric.
\begin{equation}
\kappa_{i,j} = \kappa_{-i,-j}.
\end{equation}
\end{prior}

%%%%%%%%%%%%%%%%%%%%%%%%%%%%%%%%%%%%%%%%%%%%%%%%%%%%%%%%%%%%%%%%%%%%%%%%%%%%%
\begin{prior} The density gradient should point within $\theta=\FortyFiveDeg$ of the center.
\begin{equation}
\begin{array}{lcc}
[\begin{array}{cc} i & j \end{array}] M \nabla \kappa_{i,j} &\ge& 0, \\

[\begin{array}{cc} i & j \end{array}] M^T \nabla \kappa_{i,j} &\ge& 0, \\
\end{array}
\end{equation}
where
\begin{equation}
M = \bigg[\begin{array}{lr} \cos\theta & -\sin\theta \\ 
                              \sin\theta &  \cos\theta \end{array} \bigg],
\end{equation}
\begin{equation}
\nabla\kappa_{i,j} \equiv (2a)^{-1}(\kappa_{i+1,j} - \kappa_{i-1,j} - \kappa_{i,j+1} - \kappa_{i,j-1})
\end{equation}
and $a$ is the pixel size.
\end{prior}
This complicated expression is just saying that if the density gradient of a
pixel were pointing at most $\theta$ away from the center then moving the
pixel's position by $\theta$ should align the density gradient so that it
points directly at the center.  If the gradient is greater than $\theta$ the
``$\ge$'' condition will not be satisfied.

%%%%%%%%%%%%%%%%%%%%%%%%%%%%%%%%%%%%%%%%%%%%%%%%%%%%%%%%%%%%%%%%%%%%%%%%%%%%%
\begin{prior} The density of a pixel must be no more than twice the average
density of its neighbors. 
\begin{equation}
\kappa_n \le 2 \frac{1}{N(n)}\sum_{i\in N(n)}\kappa_i, \qquad n \ne 1
\end{equation}
\end{prior}
This is a weak smoothing criterion. Normally, it is not applied to the central
pixel, which can have arbitrary density.

%%%%%%%%%%%%%%%%%%%%%%%%%%%%%%%%%%%%%%%%%%%%%%%%%%%%%%%%%%%%%%%%%%%%%%%%%%%%%
\begin{prior} The mass profile must be steeper than $r^{-s}$.  Let $R_i$ be the 
set of all pixels on a discretized ``ring'' $i$ of radius $r_{R_i}$, one pixel thick. 
The number of pixels in a ring is $|R_i|$.
Let $C_i = r_{R_i}^s / |R_i|$, then
\begin{equation}
C_i \sum_{n \in R_i} \kappa_n - C_{i+1} \sum_{n \in R_{i+1}} \kappa_n \ge 0.
\end{equation}
\end{prior}
The default radial mass profile constraint has $s=0.5$. This is intentionally
rather shallow, but as explained in \cite{2004AJ....127.2604S} this is
motivated by evidence showing that total density distribution in central
regions of ellipticals is roughly isothermal, i.e. $r^{-2}$. Furthermore, the
projected gas density in the Milky Way scales as $r^{-1.75}$
\citep{1991MNRAS.252..210B}.

\bigbreak
Again, the most important thing to realize from the constraints and the
discretized lens equation is that \emph{the constraints are all linear}. They
can therefore be solved using any number of linear programming techniques.
However, rather than find one solution, the space of all solutions is sampled
to understand the distribution.

%%%%%%%%%%%%%%%%%%%%%%%%%%%%%%%%%%%%%%%%%%%%%%%%%%%%%%%%%%%%%%%%%%%%%%%%%%%%%
\subsection{Bayesian MCMC Sampling} \label{sec:Monte-Carlo Sampling}
%%%%%%%%%%%%%%%%%%%%%%%%%%%%%%%%%%%%%%%%%%%%%%%%%%%%%%%%%%%%%%%%%%%%%%%%%%%%%

The linear equations presented in \secref{sec:Creating Models} constrain the
solution space to a convex multi-dimensional polyhedron known as a simplex. The
interior points of the simplex are solutions to the linear equations.

\pl\ samples the interior points using a Monte-Carlo Markov-Chain (MCMC)
technique.  The general technique is described in condensed matter texts
\citep{1992Binney} and Bayesian books \citep{2003Saha}. Each solution is used
to reconstruct the arrival time surface, mass density contours, \HCi, etc.

The sampling method, \AlgorithmS, relies on being able to find random vertices
of the simplex.  The current implementation uses the standard linear
programming Simplex Algorithm \citep{1963Danzig, Numerical-Recipes-1st-edition,
Introduction-to-Algorithms} to maximize a given \emph{objective function}
subject to the linear constraints that form the simplex.  The maximum is
guaranteed to be at a vertex. For the present purposes, the objective function
is chosen randomly after each iteration of \AlgorithmS, thereby producing a new
vertex each time.

\begin{description}
\item[\AlgorithmS] \emph{(Sample interior points)}
\item 1. Let $\gamma_0$ be a vertex on the simplex and $i=0$ the index of the current iteration.
\item 2. Let $\alpha_i$ be a new vertex.
\item 3. Extend a line from $\alpha_i$ through $\gamma_i$ until a constraint is reached.  
Select an interior point $\gamma_{i+1}$ uniformly from the line.
\item 4. If another model is desired, increment $i$ and go back to 2, otherwise stop.
\end{description}

Because the simplex is convex by construction of linear hyperplanes,
\AlgorithmS\ is guaranteed to return models in the solution space.  

In addition to the explicit priors of the previous section, there is also a
prior imposed by the sampling strategy itself. Although clearly well-defined,
the physical significance of this prior continues to be the subject of study.
This is not a point to be lightly dismissed since it influences the derived 
distribtion of \HC. However, the strategy cannot be arbitrary and there are
very strict requirements on the way the volume can be sampled, which are
discussed below.  Numerous tests, both in this paper and others, have shown
that the weighting can be empirically justified. The key point is that many
different models must be examined. Other modeling techniques tend to assume the
correctness of the model that is fit to the data, rather than letting the data
itself reveal the model.  To quote \cite{1997eds..proc...60B}: ``We should
still aggressively explore all other classes of models that can also fit the
observations but yet which produce disjoint estimates for the time delay. The
true uncertainty in the Hubble constant is given by the union of all of these
models.''

\AlgorithmS, in effect, puts a metric on the union.  Previous \pl\ papers implied
that the sampling of the simplex was uniform in volume, but this is not
correct, nor is it desired. The space does not have a Euclidean metric and
while it is still unclear what metric the space \emph{should} have, there are
    certain properties that an algorithm sampling the space \emph{must} have.

\begin{enumerate}
\item 
The sampling strategy must be insensitive to changes in dimensionality of the
space. In other words, increasing the number of variables (e.g. by increasing
the pixel resolution, which subdivides pixels) should not change the predicted
values of \HC.  This is not true if the solution space is uniformly sampled. As
an example of the problem, imagine a uniformly sampled right triangle where the
legs meet at the origin. The density of points projected onto one axis will be
greater towards the origin. In higher dimensions, when the points are projected
onto the same axis, the density distribution will be skewed further towards the
origin. 

\item 
The sampling strategy must be insensitive to units.  The variables that define
the solution space do not all have the same units.  Some are mass density, some
are source positions, one is \HC, etc. By simply scaling any of these units the
space is stretched or compressed. This would affect a sampling strategy
based on volume when the number of dimensions is greater than two. Whatever the
sampling prior is, it \emph{must} be insensitive to this.  
\end{enumerate}

\begin{figure}[\figplace]
\plotone{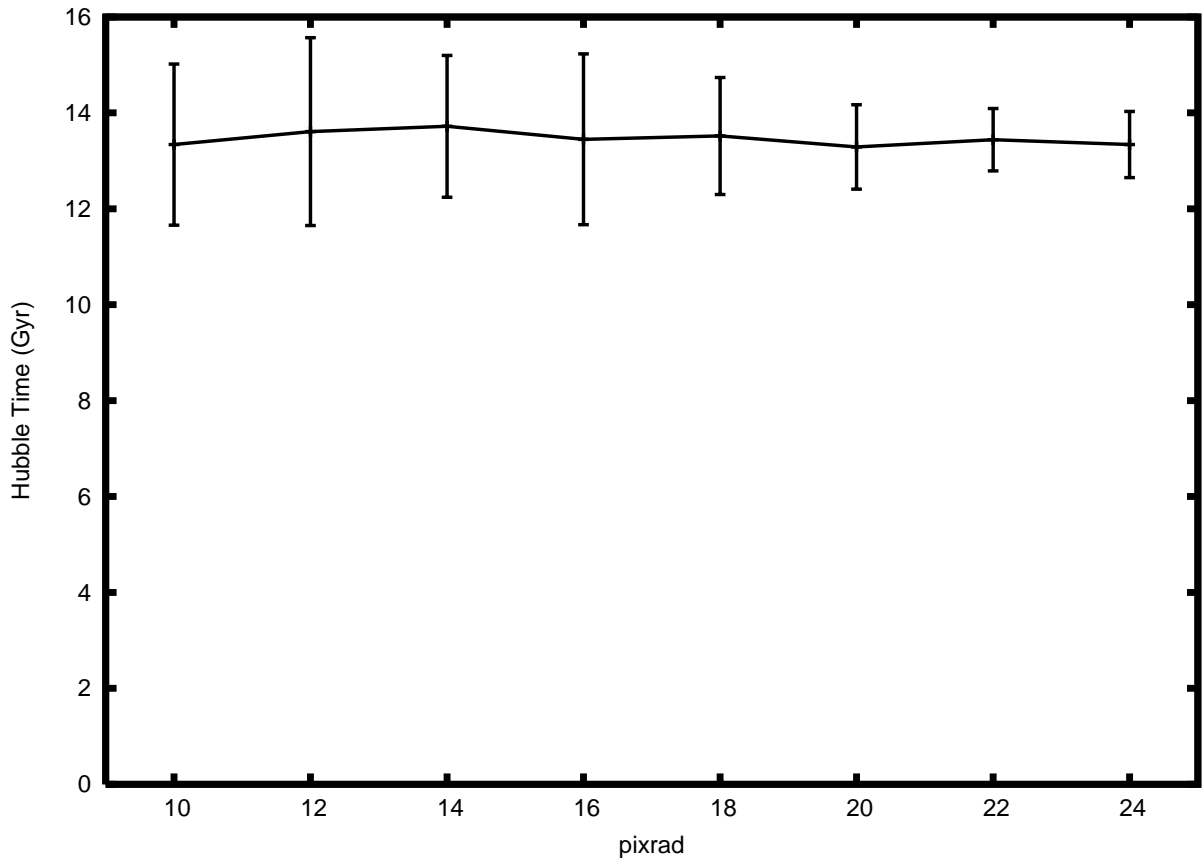}
\caption{
The predicted Hubble Time as a function of the pixel radius of the grid. The
number of variables is $O(\mathrm{{\tt pixrad}}^2)$. Error bars indicate
the $1\sigma$ deviations from the medians. Increasing the variables does not
grossly affect the median \HC, showing that condition (1) of the sampling strategy
is satisfied. A single lens, B1115+080, is being modeled. 
}
\label{fig:Converging errors}
\end{figure}

\begin{figure}[\figplace]
\plotone{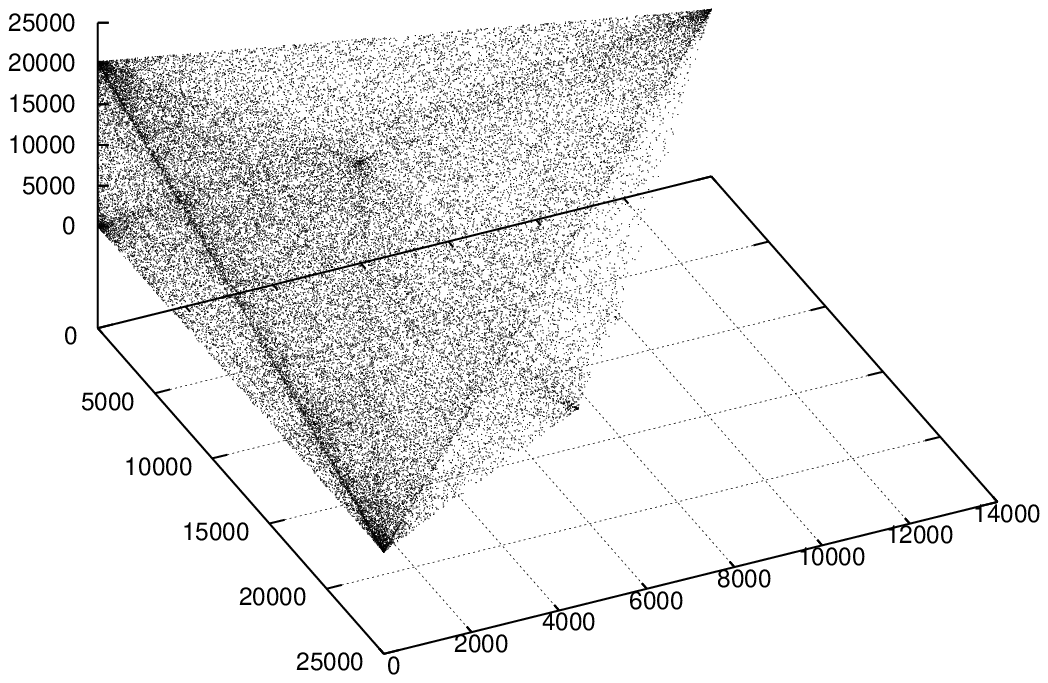} 
\caption{A three dimensional example of a sampled simplex with 50,000 points.
The overdensities clearly indicate that the volume is not uniformly sampled. 
This must be the case in order to satisfy the two conditions of the sampling strategy. }
\label{fig:3dProblem} 
\end{figure}

Both of these serious problems are solved by \AlgorithmS. The first problem is
solved because a point is chosen uniformly along the line in step 3 regardless
of the number of dimensions. One can see from \figref{fig:Converging errors}
that the predicted value of \HC\ remains quite steady even as the number of
variables is increasing.  The second problem is solved because the vertices of
the space are used to guide the sampling strategy. How the vertices are chosen
is completely independent of units. Any scaling would not affect the vertex
selection procedure. \figref{fig:3dProblem} shows a three dimensional sampled
simplex. The sampling is clearly not uniform but is insensitive to stretching
of the axes.

\AlgorithmS\ has changed slightly from older versions of \pl.  Previous
versions took a fixed number of vertex steps. The new vertex was often very
close to the old one and resulted in clumps of correlated models.  The new version
seeks out vertices further away, which reduces the problem and better samples
the interior with fewer samples.  The running time increased with this change,
but the results are more representative. Within the errors, though, old results
are still valid.

%Consider the three dimensional problem 
%\begin{equation}
%\begin{array}{rcrcrcr}
%x &+& y &+& z &\le& 25000 \nonumber \\
%0.06x &+& 0.07y &+& 0.08z &\le& 1620 \nonumber \\
%& & y &-& z &\le& 6000 \nonumber
%\end{array}
%\label{3dProblemEqs}
%\end{equation}
%shown graphically in \figref{fig:3dProblem}. 

\begin{figure}[\figplace]
\plotone{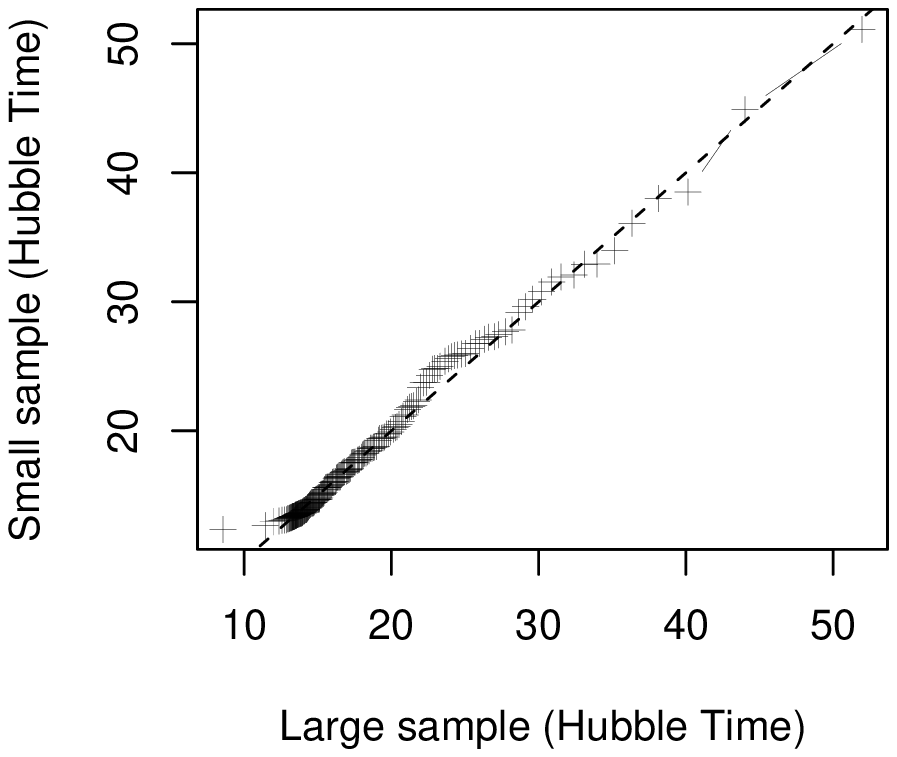}
\caption{Quantile-quantile plot comparing the distribution of a large sample of
Hubble Times to the distribution of a small sample. The points lie nearly
perfectly on the dashed line, indicating that the two samples come from the
same distribution. The tail extends off the figure because of a few extreme
outliers in the large sample. The figure was clipped for clarity.}
\label{fig:Hubble QQplot}
\end{figure}

Although \AlgorithmS\ does not sample the volume uniformly, in the limit of
infinite samples, it does have \emph{some} distribution.  But how well is that
distribution recovered with only a finite number of samples?  
To approximate the true distribution ten thousand models of the lens B1115+080 were generated.
The ``finite'' sample consisted of 200 models.  \figref{fig:Hubble QQplot} compares
the distribution of just the Hubble Time variable.  When the two samples are taken from
the same distribution, the crosses fall on the dashed line.  Even with a small
sample, the distribution is well recovered.

%%%%%%%%%%%%%%%%%%%%%%%%%%%%%%%%%%%%%%%%%%%%%%%%%%%%%%%%%%%%%%%%%%%%%%%%%%%%%%
\subsection{Technical Issues} \label{sec:Technical Issues}
%%%%%%%%%%%%%%%%%%%%%%%%%%%%%%%%%%%%%%%%%%%%%%%%%%%%%%%%%%%%%%%%%%%%%%%%%%%%%%

While \pl\ is stable, variations on sampling can introduce numerical
instability.  If a point is not chosen uniformly from the line in step 3 of
\AlgorithmS\ a numerical error in the coordinates of sampled points will grow
exponentially fast and lead to future points lying \emph{outside} the solution
space. Reprojecting a point back into the space is impractical because the
exact size and shape of the simplex is unknown and truly incalculable due to
the extraordinarily large number of dimensions and vertices. (It is worth noting
that if all the vertices could be known in advance, the Simplex Algorithm would
be unnecessary. One could simply pick a new vertex from the list.)
In the worst case, however, this error is detectable. If such an error is
detected the program will issue a message and halt.

\begin{figure}[\figplace]
\plotone{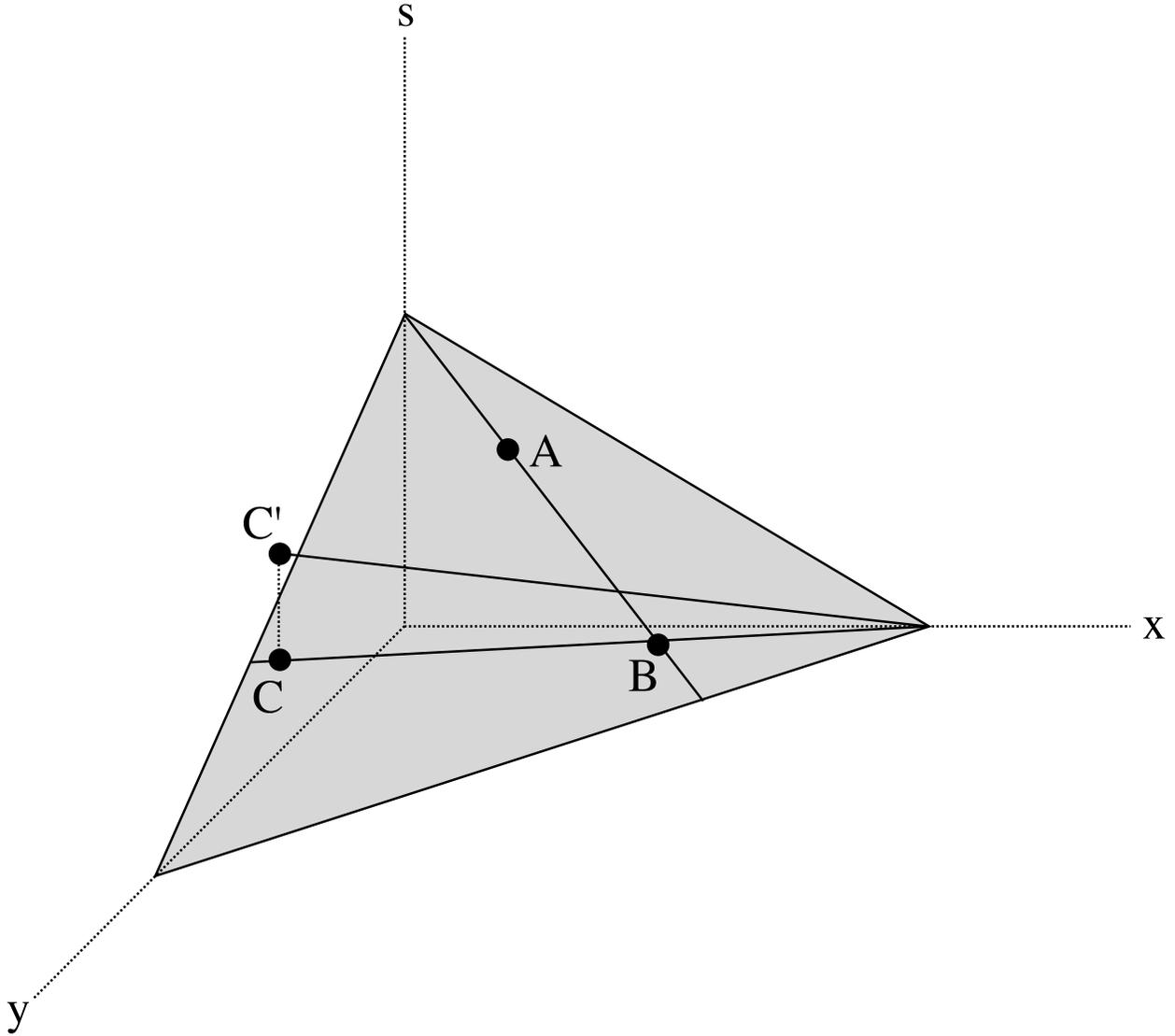}
\caption{
Example of numerical error in selecting point $C$. The $x$ and $y$ axes are the
two main variables, and $s$ is the slack variable introduction by the Simplex
linear programming algorithm. The grey region is the plane on which solutions
lie. Point $C$ lies far enough from point $B$ that numerical error is
introduced, leading to the selection of $C'$, which lies outside the grey
solution space.  Subsequent similar sampling leads to exponentially fast
growing error. The error in the diagram is exaggerated for clarity.}
\label{fig:Numerical error}
\end{figure}

The source of the error can be seen in \figref{fig:Numerical error}. The figure
has been exaggerated for clarity. Sample points are constrained to lie on the
shaded surface.  After sampling points $A$ and $B$, point $C$ is the next
intended point, but because of the limits of machine accuracy $C'$ is taken
instead.

If the problem only occurred once, the error would be below the noise in the
system, but each sample introduces more error because the next sample depends
on the position of the previous sample.

Using the notation of \AlgorithmS, the further $\gamma_{i+1}$ is chosen from
$\gamma_i$ the larger the error. This is a simple lever; the error is
proportional to $(a/b)$ where $a=\gamma_{i+1} - \alpha_i$ and $b=\gamma_i -
\alpha_i$. Successive errors are compounded over $N$ iterations:
\begin{equation}\label{eqn:eps}\epsilon = \prod_i^N (a_i / b_i). \end{equation}
Sampling uniformly along the line suppress the error because points are chosen
close to $\gamma_i$ as often as far away.
If $a_i \ge b_i$, $\epsilon$ grows without bound. If $\ln\epsilon > 0$ then
$\langle a/b \rangle \ge N$, in which case the error is reported and the program halts.

A number of technical improvements were also made to the implementation of the
Simplex Algorithm. As mentioned earlier, the Simplex Algorithm is used to find
a new vertex in \AlgorithmS\ by maximizing an \emph{objective function} subject
to the linear constraints that form the simplex. Each iteration moves to a new
vertex that increases the objective function until no further vertex can be
found.  The linear constraints are stored in a matrix called a \emph{tableau}.
The algorithm moves to the next vertex by rewriting the tableau; an operation
known as a pivot. For very large problems the pivot is the bottleneck. This
work improves the performance by parallelizing the pivot on a shared memory
machine. For even larger problems than are faced here it may be necessary to
extend this to a distributed-memory cluster of machines.

\begin{figure}[\figplace]
\plotone{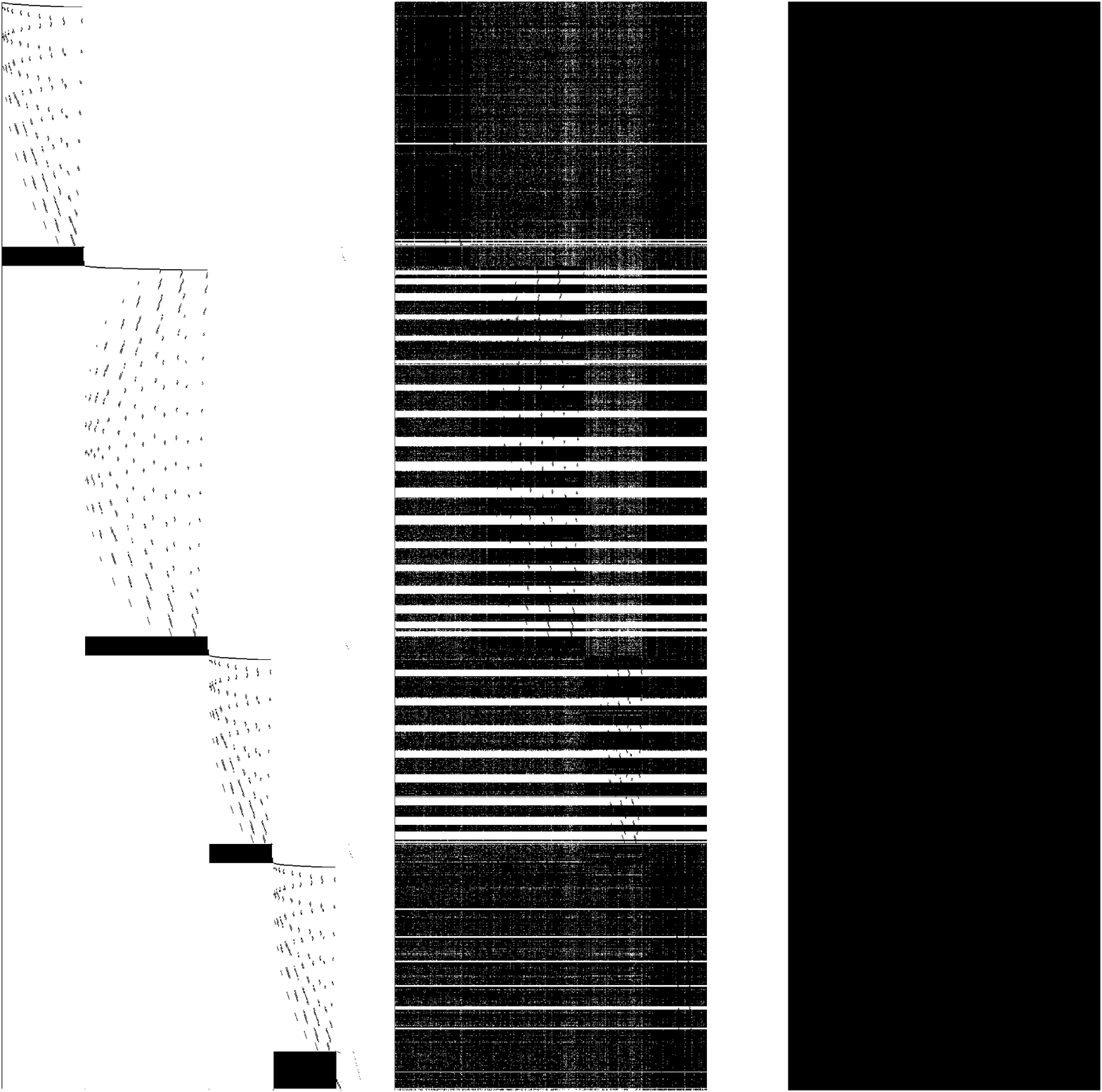}
\caption{Three simplex tableaus displayed graphically. The left image is the
tableau with the original constraints in place. Lens asymmetry can be seen in the block
that is twice as tall as the other three. The middle image is after a 
feasible solution is found. The third image is after 200 models. Black represents
non-zero values.}
\label{tableau-graphics}
\end{figure}

A further improvement was an optimization of the data structure used to store
the tableau. While the tableau is initially sparse, and previous versions of
\pl\ stored it as such, the tableau quickly becomes dense after only a few
pivots (\figref{tableau-graphics}).  Storing the tableau as a dense matrix
yields a significant performance boost.

%%%%%%%%%%%%%%%%%%%%%%%%%%%%%%%%%%%%%%%%%%%%%%%%%%%%%%%%%%%%%%%%%%%%%%%%%%%%%%
\section{Testing Hubble Time Recovery} \label{sec:Tests}
%%%%%%%%%%%%%%%%%%%%%%%%%%%%%%%%%%%%%%%%%%%%%%%%%%%%%%%%%%%%%%%%%%%%%%%%%%%%%%

How well does \AlgorithmS\ predict the Hubble Time? Two tests were performed.

\begin{figure}[\figplace]
\plottwo{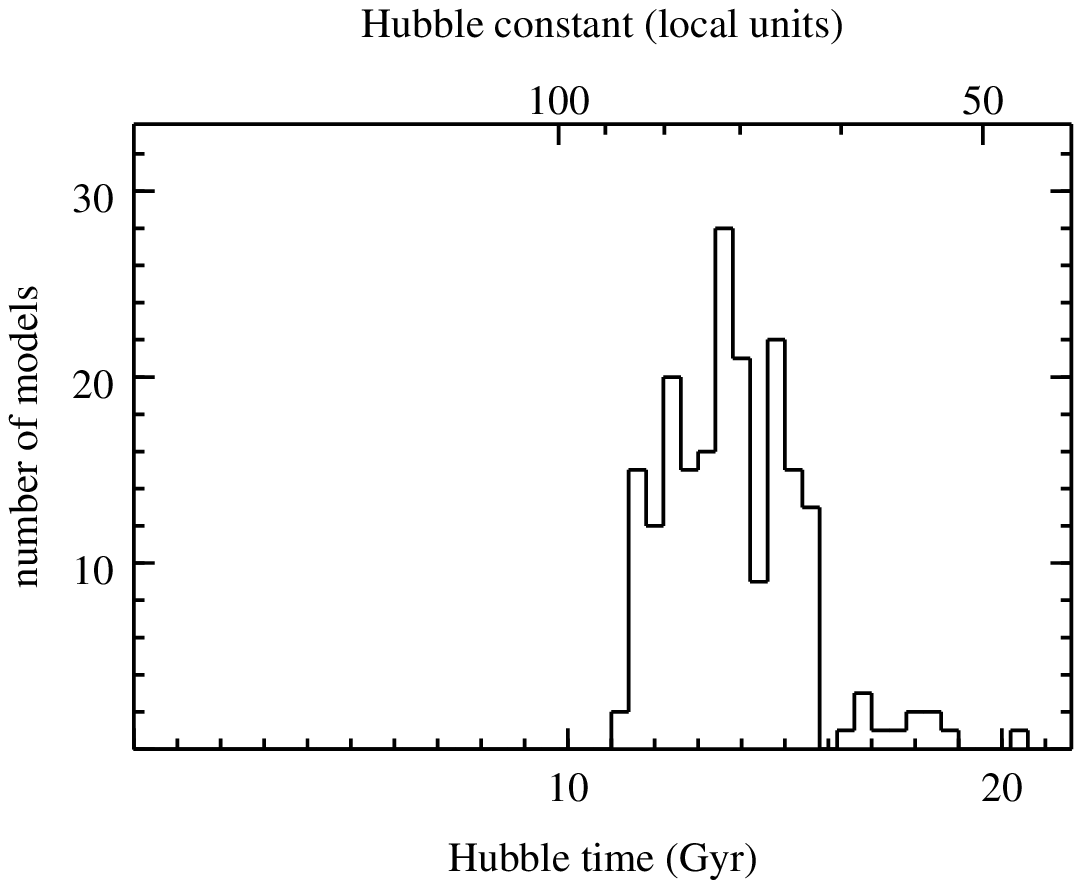}{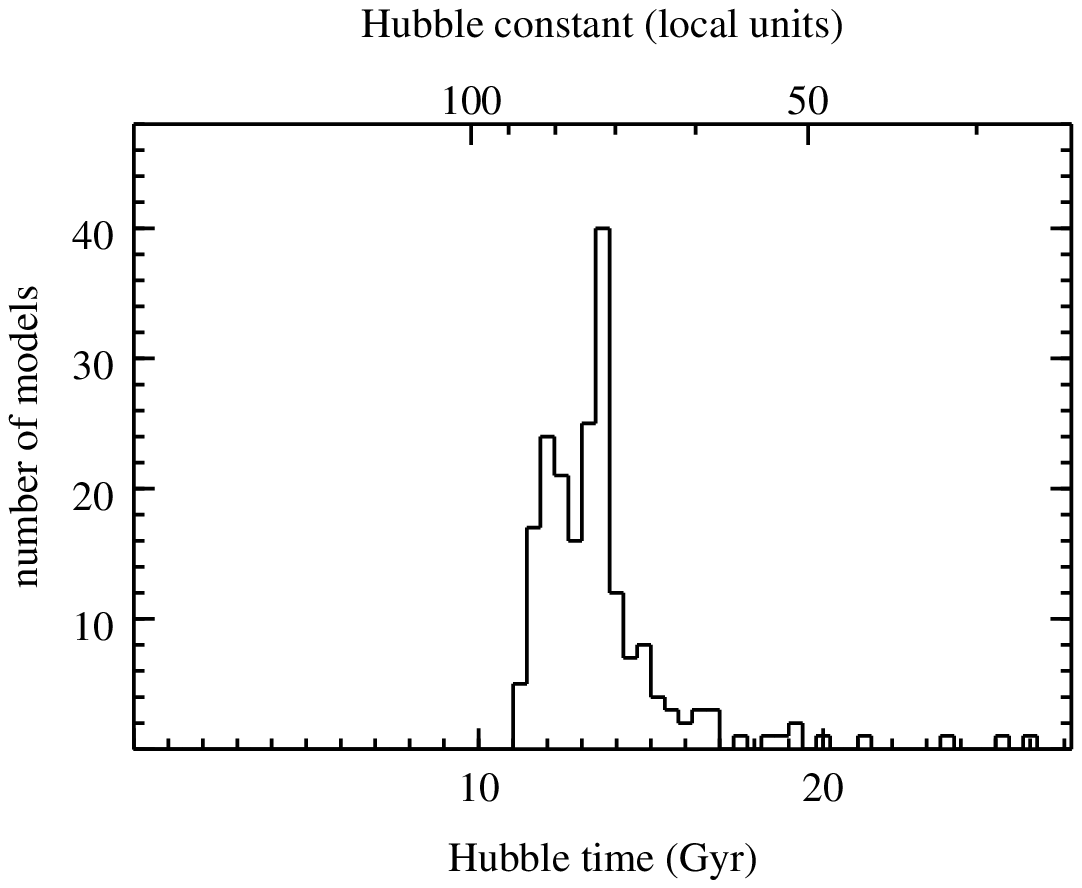}
\caption{Two tests of the program. On the left are Hubble Time values recovered
during a blind test. The lens was constructed by hand using an artificial
value of the Hubble Time (13.9 Gyr).  On the right, are time delays from a
multiply lensed simulation galaxy with $\HCi=14\mathrm{\ Gyr}$. There is a
clear peak at 13.3 Gyr. } 
\label{fig:Fake Lenses} 
\end{figure}

First, a blind test similar to that in \cite{2000AJ....119..439W}.  Four quad
lenses were crafted assuming a particular Hubble Time that was unknown to the
author. These were, in fact, the same lenses as in the aforementioned paper,
but rescaled to a Hubble Time of $13.9$ Gyr.  The time delays were perturbed
slightly to simulate errors.  The Hubble Time was recovered using \pl\ and then
the simulated Hubble Time revealed.  \figref{fig:Fake Lenses} shows the
histogram of Hubble Times from two hundred models.  \pl\ predicts \FakeResult.

Second, five lenses, three doubles and two quads, were created by ray-tracing a
galaxy from the $N$-body plus hydrodynamic simulation with \LCDMHC\ described
by \cite{2006MNRAS.366.1529M}.  The galaxy is an E1 or E2 triaxial elliptical
with about 80\% dark matter.  The histogram of Hubble Times from two hundred
models is shown on the right in \figref{fig:Fake Lenses}. There is a clear peak with
the predicted value at \SimulationResult\ with 68\% confidence. Within the
errors \pl\ successfully recovers the simulation Hubble Time.
\cite{2007arXiv0704.3267R} reconstruct the same lenses with a slightly
different prior.

%%%%%%%%%%%%%%%%%%%%%%%%%%%%%%%%%%%%%%%%%%%%%%%%%%%%%%%%%%%%%%%%%%%%%%%%%%%%%%
\section{New 11-Lens Results} \label{sec:New 11-Lens Results}
%%%%%%%%%%%%%%%%%%%%%%%%%%%%%%%%%%%%%%%%%%%%%%%%%%%%%%%%%%%%%%%%%%%%%%%%%%%%%%

\begin{figure}[\figplace]
\plotone{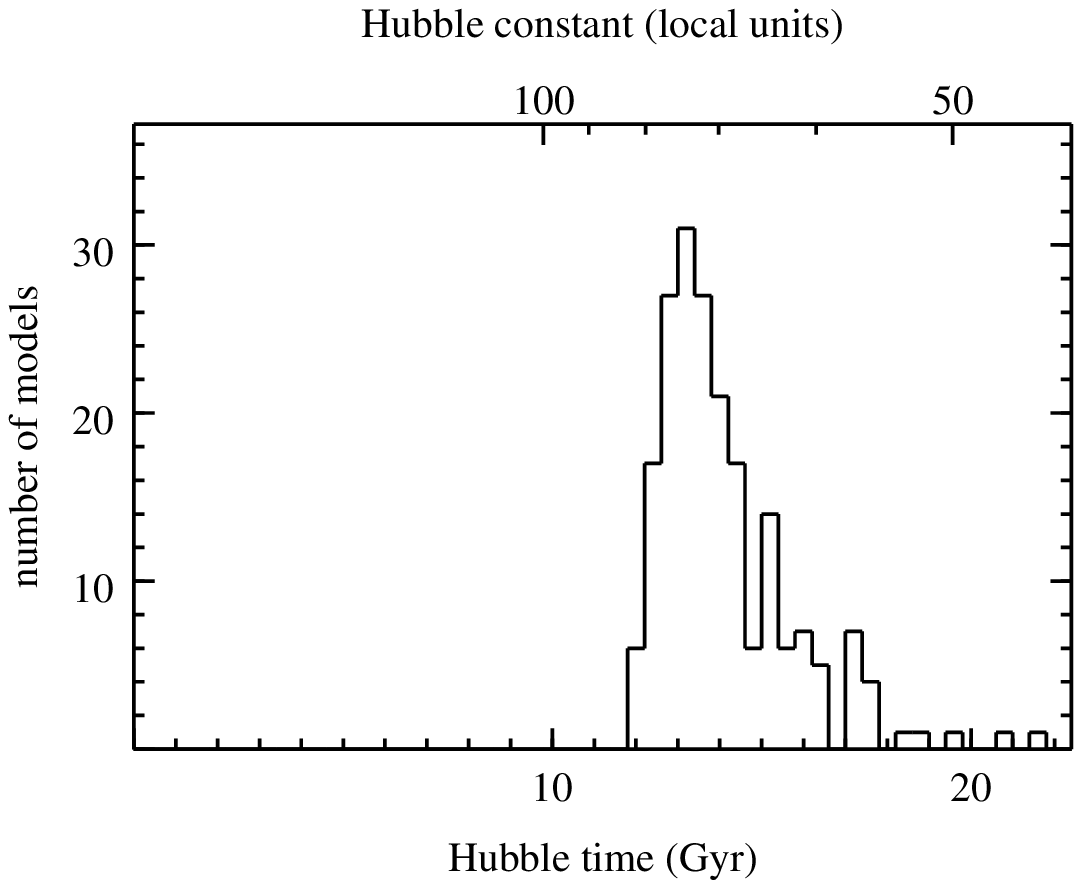}
\caption{Hubble Time values found from simultaneously modeling eleven lenses. The peak occurs at
13.7 Gyr.}
\label{fig:New Hubble Times}
\end{figure}

\begin{figure}[\figplace]
\plottwo{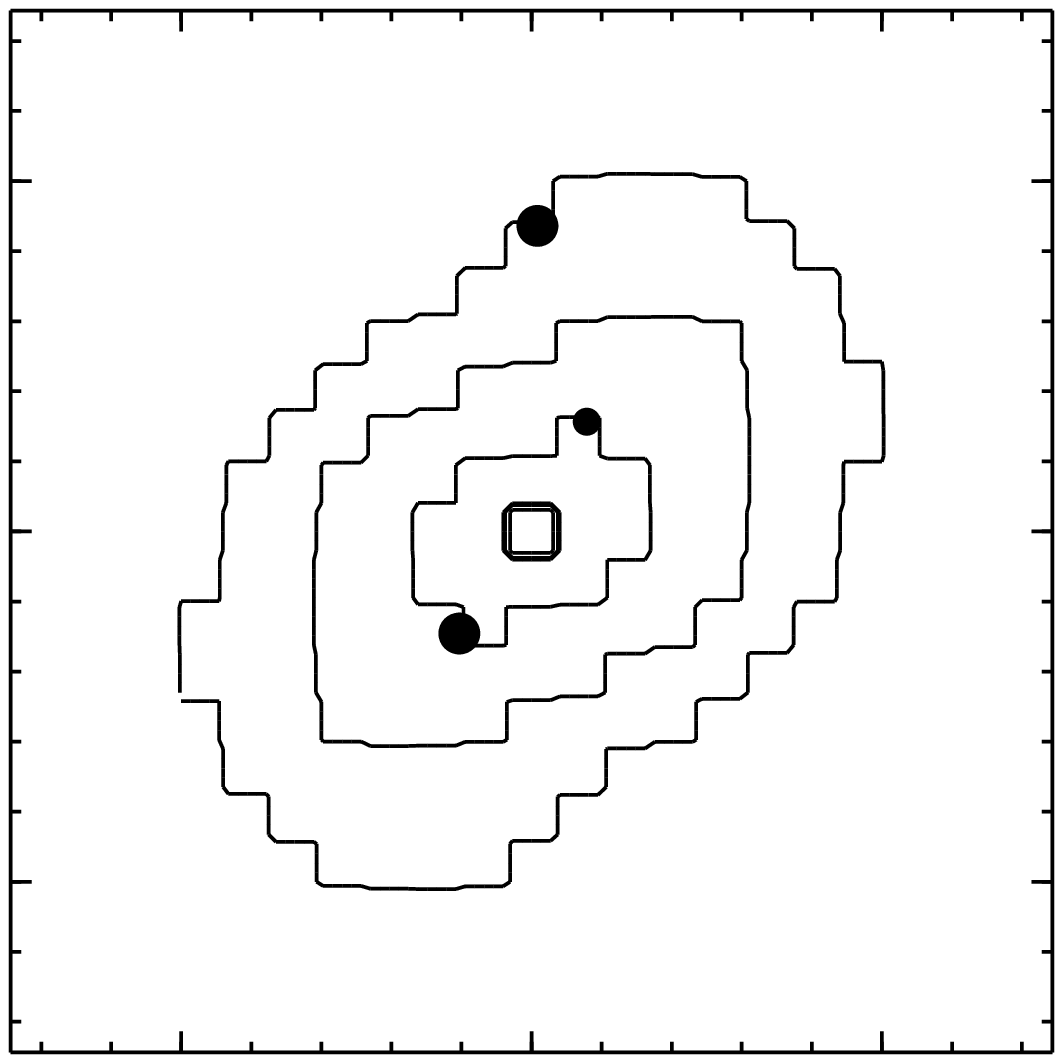}{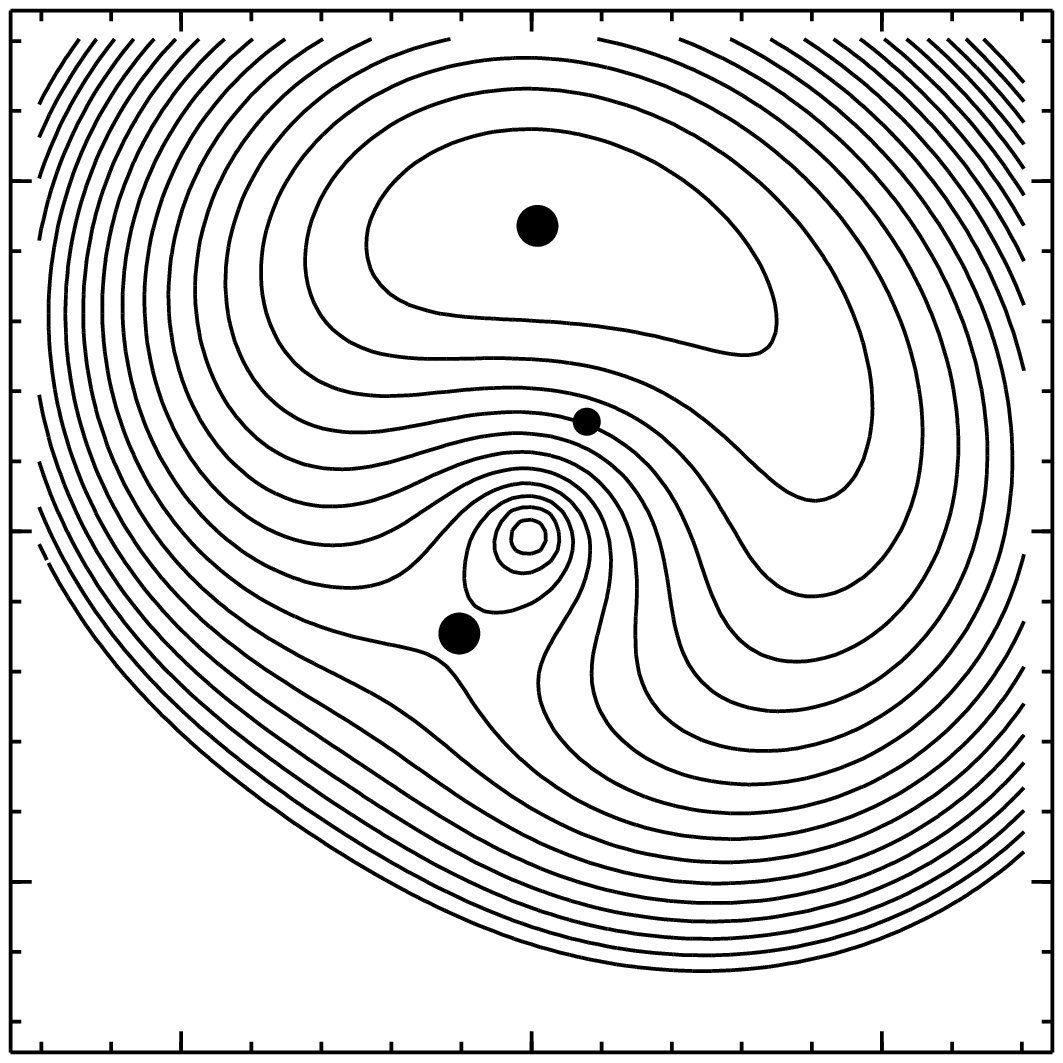}
\caption{An example of the output from \pl: The ensemble average of the mass (left) and arrival time surface (right) of J1650+4251.}
\label{fig:Models of J1650+4251}
\end{figure}

With confidence founded in the results of the last section, an ensemble of
lenses was modeled to find the true Hubble Time.
\cite{2006ApJ...650L..17S} used ten lenses%
\footnote{The ten lenses are J0911+055, B1608+656, B1115+080, B0957+561,
B1104-181, B1520+530, B2149-274, B1600+434, J0951+263, and B0218+357.}
to constrain the Hubble Time to \OldResult.
Subsequently, \cite{2007A&A...464..845V} have reported on a new time delay
measurements for J1650+4251.  Combining this new lens measurement with the ten
lenses used previous, all eleven lens were simultaneously modeled to predict
tighter bounds on the Hubble Time.
The distribution of Hubble Times is shown in \figref{fig:New Hubble Times}.  
%
%\footnote{``Eleven. Exactly. One louder.'' --Nigel Tufnel, \emph{This is Spinal Tap}.}
%
At 68\% confidence, the new predicted value is \[\NewResult\ (\NewResultLocal).\]
\figref{fig:Models of J1650+4251} shows the ensemble average of the mass and
arrival time surface for J1650+4251 as recovered by \pl. Average mass maps for
the other lenses are similar to those in \cite{2006ApJ...650L..17S}, Figure 2.

\begin{figure}[\figplace]
\plotone{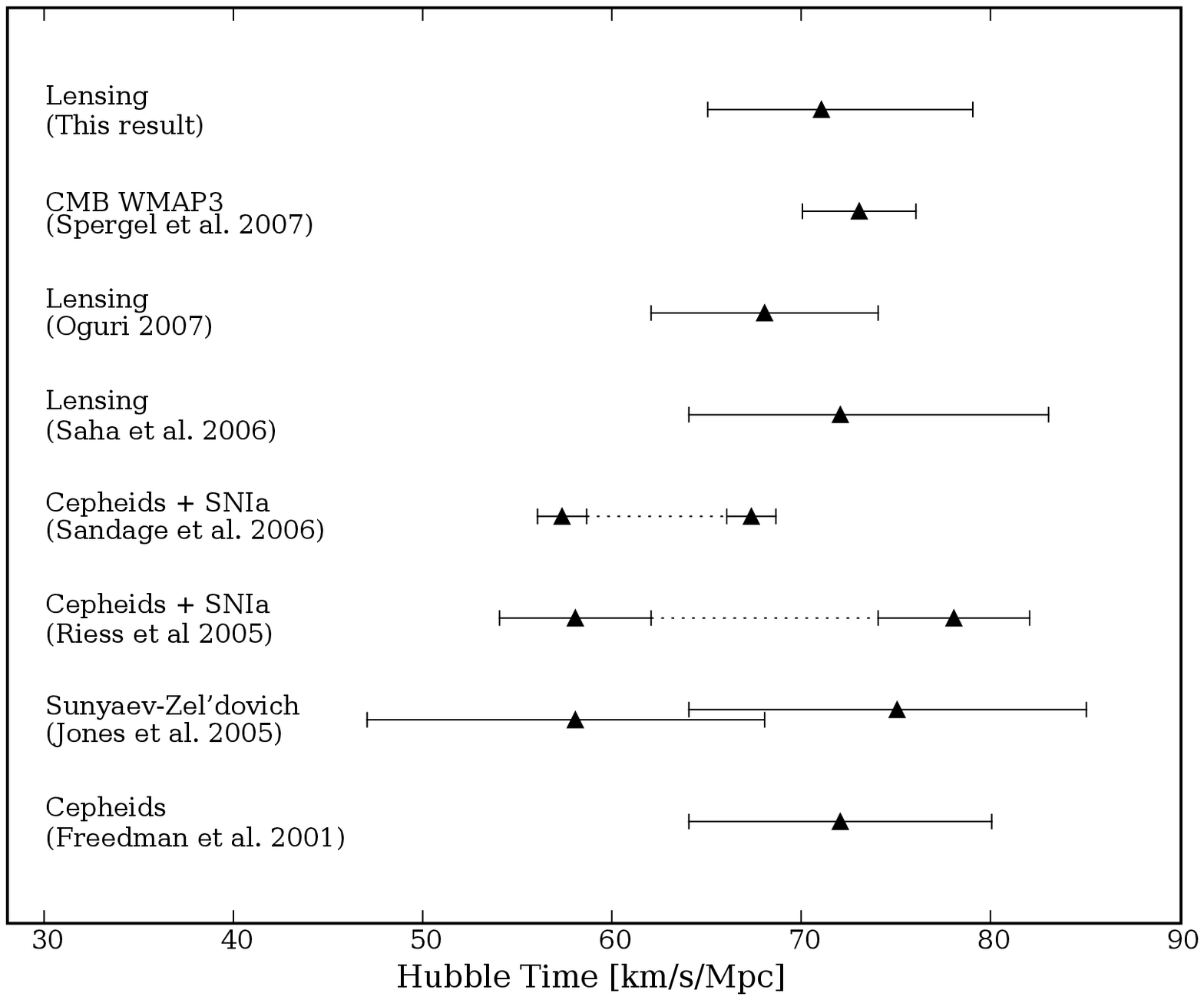}
\caption{Recent Hubble Time measurements from a variety of methods. Multiple error bars for
a single reference are present when there are systematic errors in addition to statistical.}
\label{fig:All Hubble Times}
\end{figure}

To put this into context, the results of other techniques are listed below.
The units are in \HC, which is found more often in the literature than \HCi.
The latter appears more naturally in lensing, though, hence the presentation of
the above estimates.  The first set of errors are statistical and the second
set (when applicable) are systematic. This list is summarized by the plot in
\figref{fig:All Hubble Times}.

\begin{enumerate}
\item $\HC = 73\pm3\ \kmsMpc$ from the cosmic microwave background fluctuation
spectrum \citep{2007ApJS..170..377S}. The Hubble Constant is just one value in
a multiparameter fit.

\item $\HC = 68 \pm 6\pm 8\ \kmsMpc$ using a different Monte Carlo method to
combine lenses \citep{2007ApJ...660....1O}.

\item $\HC = 62.3\pm1.3\pm5.0$ \citep{2006ApJ...653..843S} and $\HC =
73\pm4\pm5$ \citep{2005ApJ...627..579R} from Cepheid-calibrated luminosity of
Type Ia supernovae. This is independent of the global geometry.

\item $\HC = 66^{+11+9}_{-10-8}\ \kmsMpc$ from the Sunyaev-Zel'dovich effect
\citep{2005MNRAS.357..518J}. As with lensing, a global geometry is assumed and
the Hubble Time is measured.

\item $\HC = 72\pm8$ \citep{2001ApJ...553...47F} using a variety of
Cepheid-calibrated indicators. This is again, independent of the global geometry.

\end{enumerate}

In the future, better predictions may be obtained with improved priors and
tighter constraints on galaxy structure. Simply adding more lenses will also
improve the predictive power of \pl, but may not help in understanding the
different sources of degeneracies and developing better priors.

%%%%%%%%%%%%%%%%%%%%%%%%%%%%%%%%%%%%%%%%%%%%%%%%%%%%%%%%%%%%%%%%%%%%%%%%%%%%%%
\section{Summary} \label{sec:Summary}
%%%%%%%%%%%%%%%%%%%%%%%%%%%%%%%%%%%%%%%%%%%%%%%%%%%%%%%%%%%%%%%%%%%%%%%%%%%%%%

Pixelated lens reconstruction is an example of free-form modeling. Such
modeling has the advantage that one does not have to presuppose what the
important parameters might be and can let the generated models be a guide to
finding those parameters.  Free-form modeling has many applications and was
used early by Schwarzschild to show the existence of triaxial stellar systems
in equilibrium. 

Applied to gravitational lensing, the free-form models are implemented as
pixelated models whereby the mass sheet of the lens is discretized into many
small square pixels. The mass in each pixel is recovered using an MCMC
technique using linear programming to probe the solutions which reconstruct the
observed data. The software \pl\ produces an ensemble of hundreds of such
models.  The ensemble provides Bayesian statistics about the variety of possible
lens reconstructions. 

The constraints that define the mass models are explained.  The linear
constraints form a hyper-dimensional solution space from which the models are drawn.
The sampling algorithm has been improved over previous software versions and
although it was shown that the algorithm does not uniformly sample the solution
space, it is argued that this is undesirable for this problem. The
implementation was parallelized for multi-processor, shared memory machines.
Future work will include controlling numerical round-off errors that will
become significant with larger problems.

The new version of \pl\ was applied to an ensemble of eleven lenses to determine a
new value for the Hubble Time: \NewResult\ within 68\% confidence.

Further research into galaxy and cluster structure is needed to improve the
priors.  The estimates of galaxy morphology have been conservative but tighter
constraints will lead to better results. Furthermore, model ensemble building
can be applied to other areas, even to the original problems of Schwarzschild.

Pixelated lens modeling is on the cutting edge of gravitational lens research,
promising to provide great insight into the structure of galaxies, the
distribution of dark matter, and the fundamental nature of the Universe.  But
there are still many challenges both scientifically and computationally. 

%%%%%%%%%%%%%%%%%%%%%%%%%%%%%%%%%%%%%%%%%%%%%%%%%%%%%%%%%%%%%%%%%%%%%%%%%%%%%%
\section{Acknowledgments} \label{sec:Acknowledgements}
%%%%%%%%%%%%%%%%%%%%%%%%%%%%%%%%%%%%%%%%%%%%%%%%%%%%%%%%%%%%%%%%%%%%%%%%%%%%%%
I would like to extend my sincere appreciation for the help I received with
many aspects of this paper. In particular, Joachim Stadel for a critical
insight concerning the material in \secref{sec:Technical Issues}; Peter
Englmaier, Tristen Hayfield, and Justin Read for the hours spent considering
different sampling techniques; and Prasenjit Saha for patiently answering my
many lensing questions and ever so subtly nudging me to finish. I would also
like to thank the anonymous referee for useful comments and suggestions on
making the paper clearer and more concise.

%%%%%%%%%%%%%%%%%%%%%%%%%%%%%%%%%%%%%%%%%%%%%%%%%%%%%%%%%%%%%%%%%%%%%%%%%%%%%%
\bibliography{ms,apj-jour}
%%%%%%%%%%%%%%%%%%%%%%%%%%%%%%%%%%%%%%%%%%%%%%%%%%%%%%%%%%%%%%%%%%%%%%%%%%%%%%

%%%%%%%%%%%%%%%%%%%%%%%%%%%%%%%%%%%%%%%%%%%%%%%%%%%%%%%%%%%%%%%%%%%%%%%%%%%%%%
\appendix
%%%%%%%%%%%%%%%%%%%%%%%%%%%%%%%%%%%%%%%%%%%%%%%%%%%%%%%%%%%%%%%%%%%%%%%%%%%%%%

\section{\pl\ Gravitational Lens Modeling Software} \label{AppendixA}

\pl\ is freely available under the GNU General Purpose License. Source code is
naturally included. The program is cross-platform and an I/O limited version
even runs in a web browser. The version used in this paper is v1.88. For more
information, visit {\tt http://www.qgd.uzh.ch}.

Input data to \pl\ used in this paper is available with the on-line version.

\end{document}